\documentclass[%
reprint,
showpacs,preprintnumbers,
 amsmath,amssymb,
 aps,
]{revtex4-1}

\usepackage{graphicx}
\usepackage{dcolumn}
\usepackage{bm}
\usepackage{subfigure} 
\usepackage{color,soul}
\usepackage{ulem}
\usepackage{booktabs}
\usepackage{amsmath}

\begin{document}


\title{Four stages of droplet spreading on a spherical substrate and in a spherical cavity\\ -Surface tension versus line tension and viscous dissipation versus frictional dissipation-
}

\author{Masao Iwamatsu}
\affiliation{
Tokyo City University, Setagaya-ku, Tokyo 158-8557, Japan
}
\email{iwamatm@tcu.ac.jp}



\date{\today}

\begin{abstract}
The spreading of a cap-shaped spherical droplet of non-Newtonian power-law liquids on a completely wettable spherical substrate is theoretically studied.  Both convex spherical substrates and concave spherical cavities with  smooth or rough surfaces are considered.  The droplet on a rough substrate is modeled by either the Wenzel  or the Cassie-Baxter model.  The two sources of driving force of spreading by the surface-tension and the line tension are considered.  Also, the two channels of energy dissipation by the viscous dissipation within the bulk and the frictional dissipation at the contact line are considered.  A combined theory of spreading on a spherical substrate is constructed by including those four factors.  The spreading process is divided into four stages, each of which is governed by one of two driving forces and one of two dissipations.  It is found that the dynamic contact angle $\theta$ has a characteristic time ($t$) dependence at each stage.  It does not necessarily follow the standard power law $\theta \sim t^{-\alpha}$.  Instead, the relaxation can be a power-law with the exponent $\alpha$ different from that on a flat substrate, or it can be exponential or it can finish within a finite time.  Therefore, various spreading scenarios on a spherical substrate and in a spherical cavity are predicted.
\end{abstract}

\maketitle

\section{Introduction}
The wetting and spreading of a liquid droplet on a flat solid substrate has been studied theoretically as well as experimentally for more than a century, because it plays fundamental roles in many natural phenomena and industrial applications~\cite{Young1805,Gibbs1906,deGennes1985,Brochard-Wyart1992,Bonn2009}.  Although the spreading of liquid on solid substrates is a complicated phenomena where many factors come into play, the time evolution of spreading on a flat~\cite{Hoffman1975,Voinov1976,Tanner1979,deGennes1985,Brochard-Wyart1992,Seaver1994,deRuijter2000} solid substrate is usually described by amazingly simple universal power laws.  The most well-known law called Tanner's law, which describes the spreading on flat substrate, has been derived theoretically from several different approaches~\cite{Voinov1976,Tanner1979,deGennes1985} and confirmed experimentally~\cite{Tanner1979,deRuijter1999,deRuijter2000,Roques-Carmes2010,Bazazi2018,James2018} and numerically~\cite{deRuijter1999b,Milchev2002}. However, the wetting and spreading on a curved substrate~\cite{Tao2011,Eral2011,Guilizzoni2011,Extrand2012,Wu2015} and the intrusion and spreading within a curved wall of cavity~\cite{Lefevre2004,Li2008,Bormashenko2013,Iwamatsu2016d,Zhang2016,Tinti2017} attract less scientific attention even though they are also ubiquitous.  So far, most of the theoretical as well as experimental works of spreading are confined to the spreading on a flat substrate except for a few theoretical works on a spherical substrate~\cite{Iwamatsu2017a,Iwamatsu2017b,Iwamatsu2017c}.

The spreading of a cap-shaped droplet on a completely wettable spherical substrate was considered theoretically~\cite{Iwamatsu2017a,Iwamatsu2017b,Iwamatsu2017c}.  However, only the late stage of spreading was considered by employing the energy balance approach~\cite{deGennes1985,Iwamatsu2017a}, where the viscous energy dissipation within the bulk of the droplet is balanced by the driving force of spreading by the surface tension and the line tension.  Various power laws similar to Tanner's law with different spreading exponents were deduced.   It turned out that the spreading on a spherical substrate is totally different from that on a flat substrate~\cite{Iwamatsu2017a,Iwamatsu2017b,Iwamatsu2017c}. In particular, the contact-line radius shrinks and disappears on a completely wettable spherical substrate while it expands to infinity on a flat substrate.  Then, the effect of line tension~\cite{Gibbs1906,Schimmele2007,Law2017} in addition to the surface tension~\cite{Young1805} can be non-negligible on a spherical substrate for nano droplets~\cite{Iwamatsu2017a,Iwamatsu2017b,Iwamatsu2017c,Law2017} because the effect of line tension is inversely proportional to the contact-line radius.  In fact, not only the power-law spreading but also the faster exponential and the finite time spreadings have been predicted~\cite{Iwamatsu2017a, Iwamatsu2017b} when the line tension dominates over the surface tension.

However, it becomes well recognized that the spreading process is divided into several stages or regimes, each of which is characterized by different spreading exponents~\cite{deRuijter1999,Roques-Carmes2010,Bazazi2018,James2018}.  In fact, Tanner's hydrodynamic regime characterized by the viscous dissipation is believed to be preceded by the molecular-kinetic regime characterized by the frictional dissipation at the contact line~\cite{deRuijter1999,Roques-Carmes2010} and the initial inertia spreading regime~\cite{Biance2004,Bird2008,Nakamura2013}.  The molecular kinetic theory (MKT)~\cite{Blake1969,deRuijter1999b,deRuijter2000,Blake2006,Bertrand2009}, which describes the spreading at the molecular-kinetic regime characterized by the frictional dissipation becomes increasingly popular.  Recent molecular dynamics simulations of the spreading of nano droplets revealed that the friction coefficient becomes comparable to the dynamic viscosity~\cite{Zhao2017,Johansson2018}. 

The MKT uses the energy balance approach where the frictional energy dissipation at the contact line of droplet is balanced by the driving force of spreading by the surface tension.  A combined theory which includes both the viscous dissipation and the frictional dissipation was also developed~\cite{deRuijter1999,Roques-Carmes2010,deRuijter1999b}.  However, they were formulated only on a flat substrates.  Furthermore, only the surface tension is considered as the driving force of spreading.  In the present study, we attempt to develop a more general combined theory of spreading on a spherical substrate, where we include two channels of dissipation due to the viscous dissipation and the frictional dissipation.  In addition, we include two driving forces of spreading due to the surface tension and the line tension.  A combination of those four factors leads to the four stages of spreading.  Therefore, we revise and enlarge our previous study~\cite{Iwamatsu2017a,Iwamatsu2017b,Iwamatsu2017c} by including the frictional dissipation.  However, we will not consider the very initial stage~\cite{Biance2004,Bird2008,Nakamura2013} and the final relaxation stage~\cite{deRuijter1999,Roques-Carmes2010} of spreading.  Also, we will consider droplets and substrates which are smaller than the capillary length~\cite{deGennes1985} so that the effect of gravity is assumed to be negligible.

This paper is organized as follows. In Sec. II, we present the basic equation of combined theory on a 
spherical substrate based on the energy balance approach. For the sake of completeness, we will include the spreading not only on a spherical substrate but also in a spherical cavity.  We also include spreading on a rough substrate~\cite{Shuttleworth1948,McHale2004,McHale2009,Iwamatsu2017b} based on the Wenzel~\cite{Wenzel1936,Bormashenko2015} and the Cassie-Baxter~\cite{Cassie1944,Bormashenko2015} model as well.  In Sec. III, we characterize each stage of spreading by the time evolution law of dynamic contact angle.   Finally, in Sec. IV, we conclude by emphasizing the implication of our results to future experiments and simulations.

\section{\label{sec:sec2}Energy balance approach}

\begin{figure}[htbp]
\begin{center}
\includegraphics[width=0.9\linewidth]{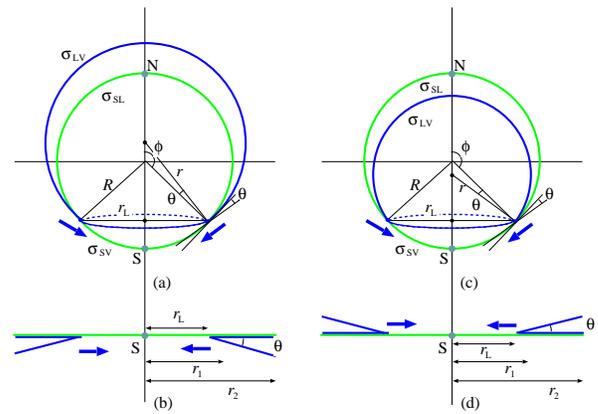}
\caption{
(a) A droplet on a convex spherical substrate spreading toward the south pole $S$ of the spherical substrate.  The three phase contact line $L$ and its radius $r_{\rm L}$ shrink towards the south pole $S$.  The meniscus of the droplet is always convex.  (b) A droplet in a concave spherical cavity.  The meniscus of the droplet is concave in the late stage.  (c) and (d) The spreading front of the droplet is approximated by the wedge on a flat surface.  Two cut-off distances $r_{1}$ and $r_{2}$ are necessary to calculate viscous dissipation.
 }
\label{fig:C1}
\end{center}
\end{figure}

We consider the late stage of spreading of a spherical cap-shaped droplet with a convex meniscus on a convex spherical substrate  [Fig.~\ref{fig:C1}(a)] and that of a lens-shaped droplet with a concave meniscus on an inner wall of concave spherical cavity  [Fig.~\ref{fig:C1}(b)].  The temporal radius of the droplet is denoted by $r=r\left(t\right)$ and that of the substrate is denoted by $R$.  We consider the late stage of spreading when the contact line locates on the lower hemisphere (lower half) of the spherical substrate so that the half of the central angle $\phi$ defined in Fig.~\ref{fig:C1} satisfies $\phi>\pi/2$~\cite{Iwamatsu2017a,Iwamatsu2017b,Iwamatsu2017c}. Since we consider the completely wettable substrate, the contact line $L$ shrinks and approaches the south pole $S$ of the spherical substrate ($\phi\rightarrow \pi$).  The radius of the contact line vanishes ($r_{\rm L}=R\sin\phi\rightarrow 0$) when the spreading is completed, and the whole substrate wall is covered by the liquid.  The novelty of this late stage spreading on a spherical geometry is that the contact line goes round the equator of the substrate and cuts in the lower hemisphere.  Then, the droplet wets the substrate almost completely but the contact line shrinks rather than expands as shown in Fig.~\ref{fig:C1}.

In the energy balance approach~\cite{deGennes1985,Iwamatsu2017a}, the spreading is assumed to be slow so that the work done by the driving force must be consumed by the energy dissipation.  The driving force $f_{\rm L}$ per unit length at the three-phase contact line $L$ is given by the sum, 
\begin{equation}
f_{\rm L}=f_{\rm s}+f_{\rm l},
\label{eq:c1}
\end{equation}
of the capillary force or surface tension force $f_{\rm s}$ of the unbalanced surface tension given by
\begin{equation}
f_{\rm s}=\sigma_{\rm LV}\left(\cos\theta_{\rm e}-\cos\theta\right),
\label{eq:c2}
\end{equation}
and the line tension force $f_{\rm l}$ given by
\begin{equation}
f_{\rm l}=-\frac{\tau}{R\tan\phi},
\label{eq:c3}
\end{equation}
where $\sigma_{\rm LV}$ is the liquid-vapor surface tension, $\tau$ is the line tension acting at the contact line, $\theta$ is the dynamic (temporal) contact angle, $\theta_{\rm e}$ is the equilibrium contact angle, which is $\theta_{\rm e}=0$ (complete wetting) and $\phi$ is the half of the central angle (Fig.~\ref{fig:C1}), which is related to the dynamic contact angle $\theta$ through
\begin{eqnarray}
\tan\phi &=& \frac{r\sin\theta}{R-r\cos\theta}\;\;\;(\mbox{sphere}),
\label{eq:c4}
\\
\tan\phi &=& -\frac{r\sin\theta}{R-r\cos\theta}\;\;\;(\mbox{cavity}).
\label{eq:c5}
\end{eqnarray}
Note that the radius $r$ of the droplet is also a function of the contact angle $\theta$ because the total volume of the droplet is preserved.  

\begin{figure}[htbp]
\begin{center}
\includegraphics[width=0.9\linewidth]{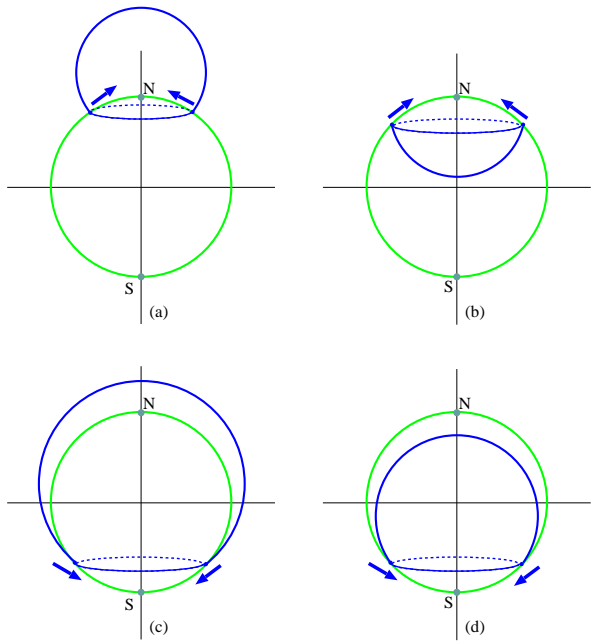}
\caption{
(a) Dewetting (complete drying) by a positive line tension on the upper hemisphere of a hydrophobic convex spherical substrate, and (b) in a hydrophobic concave spherical substrate.  (c) Spreading (complete wetting) on the lower hemisphere of a hydrophilic convex spherical substrate and (d) in a hydrophilic concave spherical cavity.   Although only the complete drying is possible by a positive line tension on a flat substrate, not only the complete drying but also the complete wetting is possible on a convex and a concave spherical substrate.  
 }
\label{fig:C2}
\end{center}
\end{figure}

When the contact line locates on the upper hemisphere of a {\it hydrophobic} substrate, a positive line tension leads to a shrinking contact line towards the north pole $N$ of the upper hemisphere [Fig.~\ref{fig:C2}(a) and (b)]. Then, the positive line tension induces {\it the complete drying (dewetting) transition}, which is similar to that on a flat substrate~\cite{Widom1995}.  On the other hand, when the contact line locates on the lower hemisphere of a {\it hydrophilic} substrate, the positive line tension leads to a shrinking contact line towards the south pole $S$ of the lower hemisphere [Fig.~\ref{fig:C2}(c) and (d)].  The positive line tension can also induce the {\it complete wetting (spreading) transition} on a spherical substrate~\cite{Iwamatsu2016a,Iwamatsu2016b,Iwamatsu2016c}, which is absent on a flat substrate because the positive line tension always resists to the spreading. Then, the complete wetting on a spherical substrate will be accelerated by the positive line tension on a spherical substrate.  In this paper, we will not consider the dewetting~\cite{Bertrand2010,Edwards2016}, but we will concentrate on the spreading towards the complete wetting state when the line tension is positive.

On the lower hemisphere of a spherical substrate, the total driving force at the contact line~\cite{Iwamatsu2017a}
\begin{equation}
f_{\rm L}=\sigma_{\rm LV}\left(\cos\theta_{\rm e}-\cos\theta\right)-\frac{\tau}{R\tan\phi}
\label{eq:c6}
\end{equation}
is always positive if $\theta>\theta_{\rm e}$ and $\phi>\pi/2$ and $\tau>0$.  The effective equilibrium contact angle $\theta_{\rm e}'$ determined from $f_{\rm L}=0$ is expressed as
\begin{equation}
\cos\theta_{\rm e}'=\cos\theta_{\rm e}-\frac{\tilde{\tau}}{\tan\phi '},
\label{eq:c7}
\end{equation}
where the half of the central angle $\phi '$ corresponds to the effective contact angle $\theta_{\rm e}'$ through Eqs. (\ref{eq:c4}) and (\ref{eq:c5}) and
\begin{equation}
\tilde{\tau}=\frac{\tau}{\sigma_{\rm LV}R}
\label{eq:c8}
\end{equation}
is the scaled line tension relative to the surface tension.  Suppose, $\tau=1$ nN~\cite{Schimmele2007,Law2017} $\sigma_{\rm LV}=73$ mNm$^{-1}$ (water), and the radius of the sphere and the cavity is $R=100$ nm, then the scaled line tension is $\tilde{\tau}\simeq 0.13$. Therefore, the effect of line tension cannot be neglected on a nano scale spherical substrate.  When the contact line locates on the lower hemisphere (Fig.~\ref{fig:C1}), the half of the central angle $\phi$ satisfies $\phi>\pi/2$, and the positive line tension $\tilde{\tau}>0$ always accelerates wetting so that the effective equilibrium contact angle $\theta_{\rm e}'$ becomes smaller than the equilibrium contact angle $\theta_{\rm e}$.  Although Eq.~(\ref{eq:c6}) is derived for the droplet on a spherical substrate~\cite{Iwamatsu2017a}, the same equation can be easily derived for the droplet in a spherical cavity~\cite{Iwamatsu2016c}.  

The above estimation $\tilde{\tau}\simeq 0.13$ seems almost an upper limit of the scaled line tension since the absolute values for line tensions reported in the literature are in the pN to $\mu$N range~\cite{Schimmele2007,Law2017,Gullemont2012}.  However, the issue of the magnitude of line tension is not conclusive.  There are some arguments that the line tension could depend on the size of the droplet and become larger for macroscopic droplets than that for nanoscale droplets~\cite{Law2017,Herminghaus2006,David2007,Heim2013}.  In fact, the line tension cannot be constant, in particular, in the last stage of spreading when $\theta\rightarrow 0$ because the effective equilibrium contact angle $\theta_{\rm e}'$ cannot be defined since the line-tension contribution in Eq.~(\ref{eq:c7}) diverges as $\theta\rightarrow 0$ or $\phi\rightarrow \pi/2$.  Then, the line tension should depend on the contact angle $\theta$ to avoid divergence~\cite{Marmur1997,Kanduc2017,Iwamatsu2018}.  We will return to this problem at the end of Sec. IIID.

\begin{figure}[htbp]
\begin{center}
\includegraphics[width=0.70\linewidth]{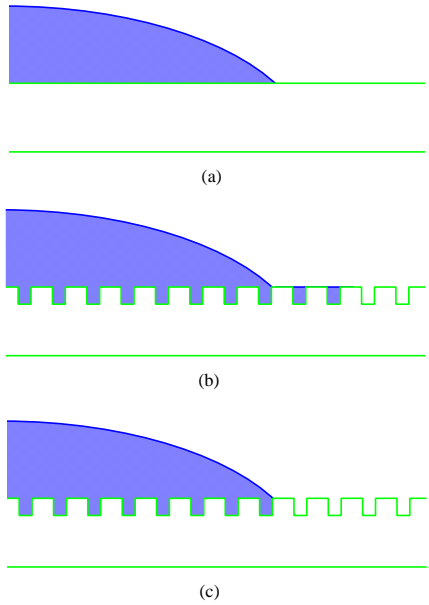}
\caption{
(a) The free droplet on a smooth substrate.  (b) The Cassie droplet of Cassie penetrating (impregnating) state, which spreads on a composite substrate composed of the solid substrate and the liquid precursor film filling the pores ahead of the contact line of the droplet. (c) The Wenzel droplet which invades the roughness of the substrate beneath the contact base of the droplet. 
}
\label{fig:C3}
\end{center}
\end{figure}

On a smooth substrate shown in Fig.~\ref{fig:C3}(a), the equilibrium contact angle $\theta_{\rm e}$ of the droplet (free droplet) is given by the Young's angle $\theta_{\rm e}=\theta_{\rm Y}$ determined from the liquid-vapor (LV) surface tension $\sigma_{\rm LV}$, the solid-liquid (SL) surface tension $\sigma_{\rm SL}$, and the solid-vapor (SV) surface tensions $\sigma_{\rm SV}$ through
\begin{equation}
\cos\theta_{\rm Y}=\frac{\sigma_{\rm SL}-\sigma_{\rm SV}}{\sigma_{\rm LV}},
\label{eq:c9}
\end{equation}
when the line tension $\tau$ is absent.  The complete wetting occurs only when $\theta_{\rm Y}=0^{\circ}$ or $\cos\theta_{\rm e}=\cos\theta_{\rm Y}=1$.  We call this completely wettable droplet on a smooth substrate the "free droplet."

It is possible to discuss the spreading on a rough textured substrate as far as the liquid volume invading into the pores of the substrate and the energy dissipation due to imbibition can be negligible~\cite{McHale2004,McHale2009,Iwamatsu2017c}.  In the complete-wetting limit, we can consider two models.  The one called Cassie penetrating (impregnating) model~\cite{Bormashenko2015,Iwamatsu2017c} shown in Fig.~\ref{fig:C3}(b) assumes that the droplet (Cassie droplet) spreads on a composite substrate made of the solid substrate and the liquid precursor film filling the pores.  The equilibrium contact angle $\theta_{\rm e}=\theta_{\rm C}$ is given by
\begin{equation}
\cos\theta_{\rm e}=\cos\theta_{\rm C}=\phi_{\rm s}\cos\theta_{\rm Y}+\left(1-\phi_{\rm s}\right),
\label{eq:c10}
\end{equation}
where the fraction $\phi_{\rm s}<1$ is defined through~\cite{McHale2009,Iwamatsu2017c} 
\begin{equation}
\phi_{\rm s}=\frac{\Delta A_{\rm non-wetted}}{\Delta A_{\rm p}},
\label{eq:c11}
\end{equation}
and $\Delta A_{\rm non-wetted}$ is a small increase in the flat-top dry area [Fig.~\ref{fig:C3}(b)] that is sampled by the advanced three-phase contact line, and $\Delta A_{\rm p}$ is the planar projection of that change~\cite{McHale2009,Iwamatsu2017c}.  The complete wetting $\theta_{\rm C}=0^{\circ}$ is realized only when the substrate itself is also completely wettable ($\theta_{\rm Y}=0^{\circ}$) or $\cos\theta_{\rm e}=\cos\theta_{\rm C}=1$.  We call this completely wettable droplet of the Cassie penetrating model the "Cassie droplet."

Another model, called the Wenzel model~\cite{Wenzel1936,McHale2004,McHale2009,Bormashenko2015,Iwamatsu2017c}, assumes that the liquid of the droplet (Wenzel droplet) invades the texture and completely wets the inside of the substrate beneath the droplet [Fig.~\ref{fig:C3}(c)].  The equilibrium contact angle $\theta_{\rm e}=\theta_{\rm W}$ is given by the Wenzel formula
\begin{equation}
\cos\theta_{\rm e}=\cos\theta_{\rm W}=r_{\rm s}\cos\theta_{\rm Y}
\label{eq:c12}
\end{equation}
with the roughness of the substrate $r_{\rm s}$ defined by~\cite{McHale2009,Iwamatsu2017c} 
\begin{equation}
r_{\rm s}=\frac{\Delta A_{\rm wetted}}{\Delta A_{\rm p}},
\label{eq:c13}
\end{equation}
where $\Delta A_{\rm wetted}$ is a small change in the wetted area that is sampled by the advanced three-phase contact line~\cite{McHale2009,Iwamatsu2017c}. In this case, the complete wetting $\cos\theta_{\rm e}=r_{\rm s}\cos\theta_{\rm Y}\ge 1$ is possible even when the substrate is not completely wettable ($\theta_{\rm Y}>0^{\circ}$) as far as $r_{\rm s}\cos\theta_{\rm Y}\ge 1$ or $r_{\rm s}>1/\cos\theta_{\rm Y}$.   We call  this completely wettable droplet of the Wenzel model the "Wenzel droplet"~\cite{Iwamatsu2017c} in this paper.

As far as the equilibrium contact angle $\theta_{\rm e}$ ($=\theta_{\rm Y}$, $\theta_{\rm C}$, $\theta_{\rm W}$) satisfies the complete wetting condition $\theta_{\rm e}=0$, the addition of the effect of line tension in Eq.~(\ref{eq:c7}) does not alter the complete wetting condition $\theta_{\rm e}'=0$.  Therefore, the line tension acts only to accelerate the spreading towards the complete wetting state.   The relative importance of the surface tension (capillary) force $f_{\rm s}$ and the line tension force $f_{\rm l}$ depends not only on the magnitude of surface tension $\sigma_{\rm LV}$ and that of line tension $\tau$ but also on the magnitude of the dynamic contact angle $\theta$ through Eqs.~(\ref{eq:c2}) and (\ref{eq:c3}).

When the contact angle $\theta$ becomes low, the surface tension force for the free and the Cassie droplet becomes
\begin{equation}
f_{\rm s} \rightarrow \sigma_{\rm LV}\frac{\theta^{2}}{2}\;\;(\mbox{free and Cassie})
\label{eq:c14}
\end{equation}
from Eq.~(\ref{eq:c2}) because $\cos\theta_{\rm e}=1$, and that for the Wenzel droplet becomes
\begin{eqnarray}
f_{\rm s}&=&\sigma_{\rm LV}\left(r_{\rm s}\cos\theta_{\rm Y}-1+\frac{\theta^{2}}{2}\right) \nonumber \\
&\rightarrow& \sigma_{\rm LV}\left(r_{\rm s}\cos\theta_{\rm Y}-1\right)\;\;\;(\mbox{Wenzel})
\label{eq:c15}
\end{eqnarray}
from Eqs.~(\ref{eq:c2}) and (\ref{eq:c12}), which is a constant acceleration force. 

The half of the central angle $\phi$ (Fig.~\ref{fig:C1}) is related to the contact angle $\theta$ through Eq.~(\ref{eq:c4}), which becomes
\begin{equation}
\tan\phi \rightarrow -\frac{r_{0}}{r_{0}-R}\theta\;\;\;(\mbox{sphere})
\label{eq:c16}
\end{equation}
as $\theta\rightarrow 0^{\circ}$, where $r_{0}$ is the droplet radius when the droplet completely encloses the spherical substrate of radius $R$($<r_{0}$), which is determined from the droplet volume $V_{0}$ through
\begin{equation}
V_{0}=\frac{4\pi}{3}\left(r_{0}^{3}-R^{3}\right)\;\;\;(\mbox{sphere}).
\label{eq:c17}
\end{equation}
Similarly, from Eq.~(\ref{eq:c5})
\begin{equation}
\tan\phi\rightarrow -\frac{r_{0}}{R-r_{0}}\theta\;\;\;(\mbox{cavity})
\label{eq:c18}
\end{equation}
as $\theta\rightarrow 0^{\circ}$ for the droplet in a spherical cavity, where $r_{0}$ is the droplet radius when the droplet completely wets the inner wall of the spherical cavity of radius $R$($>r_{0}$).  It is determined from the droplet volume $V_{0}$ through
\begin{equation}
V_{0}=\frac{4\pi}{3}\left(R^{3}-r_{0}^{3}\right)\;\;\;(\mbox{cavity}).
\label{eq:c19}
\end{equation}
Therefore, the line tension force $f_{\rm l}$ in Eq.~(\ref{eq:c3}) is asymptotically given by
\begin{equation}
f_{\rm l}=-\frac{\tau}{R\tan\phi}\rightarrow \left|\frac{r_{0}-R}{r_{0}}\right|\frac{\tau}{R\theta}
\;\;(\mbox{sphere and cavity})
\label{eq:c20}
\end{equation}
for the droplet on a spherical substrate and in a spherical cavity.  A positive line tension $\tau>0$ always accelerates the spreading ($f_{\rm l}>0$).

Hence, the line tension force $f_{\rm l}$ can be dominant in the late stage of spreading when $\theta\rightarrow 0$.  In this limit, the curvature of the contact line $1/r_{\rm L}\sim 1/\theta$ diverges and the line tension contribution to the apparent contact angle $\theta_{\rm e}'$ in Eq.~(\ref{eq:c6}) diverges.  The critical contact angle $\theta_{\rm ls}$, when the line-tension force becomes dominant, is determined from $f_{\rm l}=f_{\rm s}$, which leads to
\begin{equation}
\theta_{\rm ls}=\left|2\frac{r_{0}-R}{r_{0}}\right|^{1/3}\tilde{\tau}^{1/3}
\label{eq:c21}
\end{equation}
from Eqs.~(\ref{eq:c14}) and (\ref{eq:c20}) for the free and the Cassie droplet, and
\begin{equation}
\theta_{\rm ls}=\left(\left|\frac{r_{0}-R}{r_{0}}\right|\frac{1}{\left(r_{\rm s}\cos\theta_{\rm Y}-1\right)}\right)\tilde{\tau}
\label{eq:c22}
\end{equation}
from Eqs.~(\ref{eq:c15}) and (\ref{eq:c20}) for the Wenzel droplet when $\tau>0$.  The spreading is driven by the surface tension force $f_{\rm s}$ when the contact angle is larger than $\theta_{\rm ls}$ ($\theta>\theta_{\rm ls}$), and it is driven by the line-tension force $f_{\rm l}$ in the late stage of spreading when $\theta<\theta_{\rm ls}$.

Note that the critical contact angle $\theta_{\rm ls}$ depends on the scaled line tension $\tilde{\tau}$ defined by Eq.~(\ref{eq:c8}). Suppose $\tilde{\tau}=0.1$ and the volume of the droplet $V_{0}$ is a half of the volume of sphere ($4\pi R^{3}/6$), for example, the radius of the droplet is given by $r_{0}=\left(3/2\right)^{1/3}R$ on a spherical substrate and $r_{0}=\left(1/2\right)^{1/3}R$ in a spherical cavity.  Then, the critical contact angle $\theta_{\rm ls}$ for the free and the Cassie droplet becomes $\theta_{\rm ls}\sim 17^{\circ}$ on a spherical substrate and $\theta_{\rm ls}\sim 21^{\circ}$ in a spherical cavity from Eq.~(\ref{eq:c21}). Furthermore, if the roughness is $r_{\rm s}=1.6$ and the Young's contact angle is $\theta_{\rm Y}=20^{\circ}$ and, therefore, $r_{\rm s}\cos\theta_{\rm Y}-1\simeq 0.50$, for example, the critical contact angle for the Wenzel droplet becomes $\theta_{\rm ls}\sim 1.4^{\circ}$ on a spherical substrate and $\theta_{\rm ls}\sim 3.0^{\circ}$ in a spherical cavity from Eq.~(\ref{eq:c22}).  

In order to study the spreading of the droplet, we use the energy balance approach proposed by de Gennes~\cite{deGennes1985} instead of the hydrodynamic approach~\cite{Huh1971,Dussan1974,Snoeijer2013}.  The latter approach leads to the diverging viscous energy dissipation~\cite{Huh1971}, which can be resolved only by introducing a slipping length of fluid at the substrate~\cite{Dussan1974,Snoeijer2013}.  In contrast, the energy balance approach is more advantageous as it simplifies the analysis without the detailed knowledge of flow field within the droplet.  However, the energy balance approach is also suffered by the diverging viscous dissipation, which can be resolved by the microscopic cut off distance~\cite{deGennes1985}. It can be easily extended to include the frictional dissipation at the contact line~\cite{deRuijter1999b,Blake2006} and has been routinely used to analyze various spreading phenomena~\cite{deGennes1985,Brochard-Wyart1992,deRuijter1999,Iwamatsu2017a,Iwamatsu2017b,Iwamatsu2017c,Wang2007,Liang2012}.

In the energy balance approach, the work $2\pi r_{\rm L}f_{\rm L}U$ done by the driving force $f_{\rm L}$ at the contact line $L$ with the radius $r_{\rm L}$ of the droplet and the spreading velocity $U$ must be balanced by various energy dissipations.  In order to make our discussion as general as possible, we will consider not only the usual Newtonian liquids but also the non-Newtonian power-law liquids, whose apparent viscosity $\mu$ depends on the shear rate $\dot{\gamma}$ through
\begin{equation}
\mu=\kappa \dot{\gamma}^{n-1},
\label{eq:c23}
\end{equation}
where $\kappa$ is a consistency coefficient~\cite{Wang2007,Liang2012,Iwamatsu2017b}. The power-law exponent $n$ characterizes the non-Newtonian liquids.  When $n>1$, the liquid is called shear thickening.  When $n<1$, it is called shear thinning.  The usual Newtonian liquids correspond to $n=1$.

Similar to the driving force of spreading in Eq.~(\ref{eq:c1}), the energy dissipation $\dot{\Sigma}$ is also divided into two factors:
\begin{equation}
\dot{\Sigma}=\dot{\Sigma}_{\rm v}+\dot{\Sigma}_{\rm f},
\label{eq:c24}
\end{equation}
where $\dot{\Sigma}_{\rm v}$ is the viscous dissipation due to the flow induced within the bulk of the droplet, and $\dot{\Sigma}_{\rm f}$ is the dissipation due to the friction at the contact line.  This dissipation $\dot{\Sigma}$ in Eq.~(\ref{eq:c24}) must be compensated by the work done by the driving force $f_{\rm L}$ at the contact line given in Eq.~(\ref{eq:c1}) when the spreading velocity $U$ is low.  Then the energy balance achieves the energy balance equation,
\begin{equation}
2\pi r_{\rm L}f_{\rm L}U=\dot{\Sigma}.
\label{eq:c25}
\end{equation}
Note that the radius of the contact line shrinks ($r_{\rm L}\rightarrow 0$) and $\theta\rightarrow 0^{\circ}$ as the spreading proceeds.

In this work we neglect the liquid volume within the texture of rough substrates, and we will only consider the viscous dissipation $\dot{\Sigma}_{\rm v}$ within the droplet and neglect the dissipation due to the imbibition~\cite{Ishino2007,Grewal2015}.  Now, the spreading droplet on a spherical substrate (Fig.~\ref{fig:C1}(a)) and that in a spherical cavity (Fig.~\ref{fig:C1}(b)) can be modeled by a shrinking crater with a wedge-shaped meniscus (Fig.~\ref{fig:C1}(c), (d)).  Then the viscous dissipation $\dot{\Sigma}_{\rm v}$ within the shrinking crater is given by~\cite{Iwamatsu2017b}
\begin{equation}
\dot{\Sigma}_{\rm v}=2\pi\lambda\left(\frac{2n+1}{n}\right)^{n}\frac{\kappa U^{n+1}}{\theta^{n}}r_{\rm L}^{2-n},
\label{eq:c26}
\end{equation}
where the coefficient $\lambda$ is determined from the cut-off length $r_{1}$ and $r_{2}$ (Fig.~\ref{fig:C1}(c),(d)) of the wedge~\cite{Iwamatsu2017b}.

In addition to the viscous dissipation in the bulk $\dot{\Sigma}_{\rm v}$ given by Eq.~(\ref{eq:c26}), the friction force $\zeta U$ (per unit length) at the contact line, where $\zeta$ is a friction coefficient, gives the additional dissipation, which is written as
\begin{equation}
\dot{\Sigma}_{\rm f}=2\pi r_{\rm L}\zeta U^{2},
\label{eq:c27}
\end{equation}
where the friction coefficient is given by~\cite{deRuijter1999b,Bertrand2009}
\begin{equation}
\zeta=\frac{N_{\rm s}kT}{\nu d}
\label{eq:c28}
\end{equation}
from the molecular kinetic theory (MKT)~\cite{Blake1969,deRuijter1999b,Blake2006}, where $N_{\rm s}$ is the number of surface sites for liquid molecules per unit area of the substrate, $kT$ is the thermal energy at the temperature $T$, $d$ is the characteristic length of the displacement of liquid molecules by jumping, and $\nu$ is the frequency of molecular displacement~\cite{deRuijter1999b}.

Since the contact line shrinks ($r_{\rm L}\rightarrow 0$) as $\theta\rightarrow 0$, we consider the energy dissipation per unit length of the contact line instead of the total dissipation.  The viscous dissipation and the frictional dissipation per unit length are given respectively by
\begin{equation}
\frac{\dot{\Sigma}_{\rm v}}{2\pi r_{\rm L}}=\lambda\left(\frac{2n+1}{n}\right)^{n}\left|\frac{r_{0}}{r_{0}-R}R\right|^{1-n}\kappa U^{n+1}\theta^{1-2n},
\label{eq:c29}
\end{equation}
and
\begin{equation}
\frac{\dot{\Sigma}_{\rm f}}{2\pi r_{\rm L}}=\zeta U^{2},
\label{eq:c30}
\end{equation}
where we have used
\begin{equation}
r_{\rm L}=R\sin\phi\rightarrow \left|\frac{r_{0}}{r_{0}-R}R\right|\theta
\label{eq:c31}
\end{equation}
for the late stage of spreading from Eqs.~(\ref{eq:c16}) and (\ref{eq:c18}).  

When $n>1/2$ (shear thickening and shear thinning including Newtonian liquids), the viscous dissipation  $\dot{\Sigma}_{\rm v}$ in Eq.~(\ref{eq:c29}) diverges and becomes a main dissipation channel in Eq.~(\ref{eq:c24}) in the late stage of spreading when $\theta\rightarrow 0$.  The critical contact angle $\theta_{\rm ls}$, when the viscous dissipation dominates over the frictional dissipation, is determined from $\dot{\Sigma}_{\rm v}=\dot{\Sigma}_{\rm f}$, which leads to
\begin{eqnarray}
\theta_{\rm vf} &=& \left[\frac{\kappa\lambda}{\zeta}\left(\frac{2n+1}{n}\right)^{n}\left|\frac{r_{0}}{r_{0}-R}R\right|^{1-n} U^{n-1} \right]^{\frac{1}{2n-1}},
\nonumber \\
&&\;\;\;n> \frac{1}{2}.
\label{eq:c32}
\end{eqnarray}
On the other hand, when $n\le 1/2$ (shear thinning liquids), the viscous dissipation in Eq.~(\ref{eq:c29}) vanishes as $\theta\rightarrow 0$.  Then the frictional dissipation $\dot{\Sigma}_{\rm f}$ in Eq.~(\ref{eq:c30}) is always dominant and the critical angle $\theta_{\rm vf}$ is absent.

For Newtonian liquids with $n=1$ ($>1/2$), for example, the friction coefficient $\zeta$ is always larger than the viscosity $\mu=\kappa$ and is given by $\zeta/\kappa\approx 100-300$~\cite{deRuijter1999,Blake2006}.   Suppose $\lambda=10$~\cite{Iwamatsu2017c},  the critical contact angle becomes $\theta_{\rm vf}\approx 10^{\circ}-30^{\circ}$.  The viscous dissipation is dominant in the late stage of spreading when $\theta<\theta_{\rm vf}$, while the frictional dissipation is dominant when $\theta>\theta_{\rm vf}$.  Although the critical contact angle $\theta_{\rm ls}$ can be macroscopic only when the size of the substrate $R$ is nanoscale, the critical contact angle $\theta_{\rm vf}$ can be always macroscopic.  Therefore, it would be realistic to assume that $\theta_{\rm ls}<\theta_{\rm vf}$. 

The energy balance equation in Eq.~(\ref{eq:c25}) for the free and the Cassie droplet is written as
\begin{eqnarray}
&&\sigma_{\rm LV}\left(\frac{1}{2}\theta^{2}+\left|\frac{r_{0}-R}{r_{0}}\right|\frac{\tilde{\tau}}{\theta}\right)
\nonumber \\
&&=\lambda\left(\frac{2n+1}{n}\right)^{n}\left|\frac{r_{0}}{r_{0}-R}R\right|^{1-n}\kappa U^{n}\theta^{1-2n}+\zeta U
\label{eq:c33}
\end{eqnarray}
from Eqs.~(\ref{eq:c14}), (\ref{eq:c20}), (\ref{eq:c29}), and (\ref{eq:c30}) when the dynamic contact angle $\theta$ is low.  

Similarly, the energy balance equation for the Wenzel droplet on a rough substrate is given by
\begin{eqnarray}
&&\sigma_{\rm LV}\left(r_{\rm s}\cos\theta_{\rm Y}-1+\left|\frac{r_{0}-R}{r_{0}}\right|\frac{\tilde{\tau}}{\theta}\right)
\nonumber \\
&&=\lambda\left(\frac{2n+1}{n}\right)^{n}\left|\frac{r_{0}}{r_{0}-R}R\right|^{1-n}\kappa U^{n}\theta^{1-2n}+\zeta U
\label{eq:c34}
\end{eqnarray}
from Eqs.~(\ref{eq:c15}), (\ref{eq:c20}), (\ref{eq:c29}), and (\ref{eq:c30}).  

For Newtonian liquids with $n=1$, in particular, Eq.~(\ref{eq:c33}) for the free and the Caasie droplet is written as
\begin{equation}
U = \frac{\sigma_{\rm LV}\left(\frac{1}{2}\theta^{2}+\left|\frac{r_{0}-R}{r_{0}}\right|\frac{\tilde{\tau}}{\theta}\right)}{\zeta + \frac{3\lambda\kappa}{\theta}}.
\label{eq:c35}
\end{equation}
Similarly, Eq.~(\ref{eq:c34}) for the Wenzel droplet is written as
\begin{equation}
U = \frac{\sigma_{\rm LV}\left(r_{\rm s}\cos\theta_{\rm Y}-1+\left|\frac{r_{0}-R}{r_{0}}\right|\frac{\tilde{\tau}}{\theta}\right)}{\zeta + \frac{3\lambda\kappa}{\theta}},
\label{eq:c36}
\end{equation}
which have to be solved numerically~\cite{deRuijter1999,Blake2006}.

\begin{figure}[htbp]
\begin{center}
\includegraphics[width=0.9\linewidth]{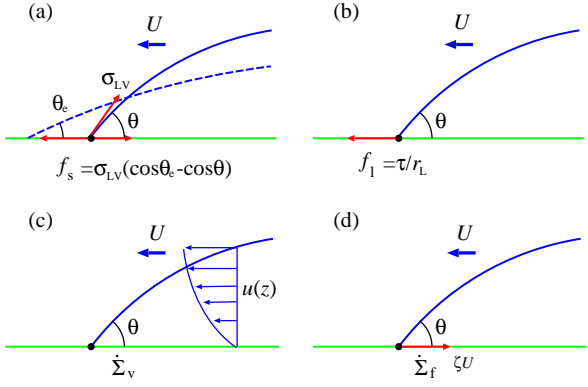}
\caption{
(a) Surface-tension force $f_{\rm s}$ and (b) line-tension force $f_{\rm l}$ are the two driving forces of spreading with the velocity $U$.  (c) Viscous dissipation $\dot{\Sigma}_{\rm v}$ due to the induced flow $u(z)$ within the bulk  and (d) frictional dissipation $\dot{\Sigma}_{\rm f}$ by the frictional force $\zeta U$ at the contact line  are the two channels of energy dissipation.  Energy balance is achieved by equating the works done by the two driving forces and the energy dissipation by the two channels, which is given by Eqs.~(\ref{eq:c33}) for the free and the Cassie droplets, and (\ref{eq:c34}) for the Wenzel droplets.
 }
\label{fig:C4}
\end{center}
\end{figure}

Equations (\ref{eq:c33}) and (\ref{eq:c34}) are the most basic equation of the combined theory of spreading for non-Newtonian droplets on a spherical substrate and in a spherical cavity, which include the surface tension and the line tension as well as the viscous dissipation and the frictional dissipation.  The left-hand side of Eqs.~(\ref{eq:c33}) and (\ref{eq:c34}) consists of the work done by the surface tension force ($f_{\rm s}$) and the line-tension force ($f_{\rm l}$), and the right-hand side consists of the dissipation due to the viscosity ($\dot{\Sigma}_{\rm v}$) within the bulk and due to the friction at the contact line ($\dot{\Sigma}_{\rm f}$).  Figure \ref{fig:C4} schematically shows the two origins of the driving force of spreading [Fig.~\ref{fig:C4}(a), (b)] and two channels of the energy dissipation [Fig.~\ref{fig:C4}(c), (d)].

In the late stage of spreading when $\theta<\theta_{\rm ls}$, the line-tension force [Fig.~\ref{fig:C4}(b)] dominates over the surface-tension force [Fig.~\ref{fig:C4}(a)) in the left hand side of Eqs.~(\ref{eq:c33}) and (\ref{eq:c34}).  Similarly, when $\theta<\theta_{\rm vf}$, the viscous dissipation [Fig.~\ref{fig:C4}(c)] dominates over the frictional dissipation [Fig.~\ref{fig:C4}(d)] in the right hand side of Eqs.~(\ref{eq:c33}) and (\ref{eq:c34}) if $n>1/2$. When $n\le1/2$ (shear thinning liquid), the frictional dissipation is always dominant.   Therefore, we can divide the spreading into four stages characterized by two critical contact angles $\theta_{\rm ls}$ and $\theta_{\rm vf}$ when $n>1/2$ and two stages characterized by $\theta_{\rm ls}$ when $n\le 1/2$, which will be discussed in detail in the next section.

\section{\label{sec:sec3}Four stages of spreading on a spherical substrate and in a spherical cavity}

It is possible to study the spreading by numerically solving the evolution equation in Eqs.~(\ref{eq:c33}) and (\ref{eq:c34}) of the combined theory~\cite{deRuijter1999,Blake2006}.  However, it will be more instructive to divide the spreading process into several stages, each of which is characterized by a scaling law and the corresponding scaling exponent~\cite{deRuijter1999,Roques-Carmes2010,Bazazi2018,James2018}.  

In the early stage of spreading (stage I) when $\theta>\theta_{\rm ls}$ and $\theta>\theta_{\rm vf}$ if $n>1/2$, and $\theta>\theta_{\rm ls}$ if $n\le 1/2$, the main driving force of spreading is the surface-tension (capillary) force ($f_{\rm s}$) and the main dissipation channel is the frictional dissipation ($\dot{\Sigma}_{\rm f}$).   

In the second stage (stage II) when $\theta_{\rm ls}>\theta>\theta_{\rm vf}$ if $n>1/2$ or $\theta_{\rm ls}>\theta$ if $n\le 1/2$, the main driving force of spreading 
becomes the line-tension force ($f_{\rm l}$) but the main dissipation channel remains the frictional dissipation.  This is the final stage of spreading of the non-Newtonian shear-thinning droplet with $n\le 1/2$.  This stage may not appear even if when $\theta_{\rm vf}>\theta_{\rm ls}$.  

In the third intermediate stage (stage III) when $\theta_{\rm vf}>\theta>\theta_{\rm ls}$, the main driving force of spreading is the surface-tension (capillary) force ($f_{\rm s}$) and the main dissipation channel becomes the viscous dissipation ($\dot{\Sigma}_{\rm v}$).  This stage III and the next fourth stage (stage IV) exist only when $n>1/2$.  However, this stage may not appear when $\theta_{\rm ls}<\theta_{\rm vf}$.  Therefore, the intermediate stage will be either II or III depending on the relative magnitudes of $\theta_{\rm vf}$ and $\theta_{\rm ls}$.
Then, the order of the stage will be  I$\rightarrow$III$\rightarrow$IV if $\theta_{\rm vf}>\theta_{\rm ls}$ and I$\rightarrow$II$\rightarrow$IV if $\theta_{\rm vf}<\theta_{\rm ls}$.

In the fourth stage (stage IV) when $\theta<\theta_{\rm ls}$ and $\theta<\theta_{\rm vf}$, the main driving force of spreading is the line-tension force ($f_{\rm l}$) and the main dissipation channel becomes the viscous dissipation ($\dot{\Sigma}_{\rm v}$)  This is the last stage of spreading when $n>1/2$ and was the subject of our previous studies~\cite{Iwamatsu2017a,Iwamatsu2017b,Iwamatsu2017c}.  In the followings, we will summarize the time evolution of the contact angle at each stage.

\subsection{Stage I: $\theta>\theta_{\rm ls}$ and $\theta>\theta_{\rm vf}$ }

In stage I, the main driving force of spreading is the surface tension force $f_{\rm s}$ in Eq.~(\ref{eq:c2}) and the main dissipation channel is the frictional dissipation $\dot{\Sigma}_{\rm f}$ in Eq.~(\ref{eq:c27}).  Note that we will not consider the precedent very first stage of spreading where the inertia effect plays a substantial role~\cite{Biance2004,Bird2008,Nakamura2013}.  The energy dissipation by the contact-line friction has been considered on a flat substrate using the so-called molecular kinetic theory (MKT)~\cite{Blake1969,deRuijter1999b,deRuijter2000,Blake2006,Bertrand2009,Zhao2017}.  Here, we extend this MKT to the convex and the concave spherical substrate.  The energy balance equation for the free and the Cassie droplet is given by
\begin{equation}
\sigma_{\rm LV}\frac{1}{2}\theta^{2}=\zeta U
\label{eq:c37}
\end{equation}
from Eq.~(\ref{eq:c33}) for a convex and a concave spherical substrate when the contact angle $\theta$ is low.  Then, the spreading velocity $U$ is proportional to the square of the contact angle, $\theta$:
\begin{equation}
U = K_{\rm I} \theta^{2},
\label{eq:c38}
\end{equation}
where 
\begin{equation}
K_{\rm I}=\frac{\sigma_{\rm LV}}{2\zeta}
\label{eq:c39}
\end{equation}
is a proportionality constant.

When the contact angle $\theta$ and, therefore, $\pi-\phi$, is low, the spreading velocity $U$ on a spherical substrate and in a spherical cavity is given by~\cite{Iwamatsu2017b}
\begin{equation}
U=\frac{d}{dt}R\phi=-\left|\frac{r_{0}}{r_{0}-R}R\right|\dot{\theta}
\label{eq:c40}
\end{equation}
from Eqs.~(\ref{eq:c16}) and (\ref{eq:c18}), where $\dot{\theta}=d\theta/dt<0$.  Then, Eq.~(\ref{eq:c37}) is written as
\begin{equation}
\theta^{2}=-\Gamma_{\rm I}\dot{\theta},
\label{eq:c41}
\end{equation}
where
\begin{equation}
\Gamma_{\rm I}=\frac{2\zeta}{\sigma_{\rm LV}}\left|\frac{r_{0}}{r_{0}-R}R\right|
\label{eq:c42}
\end{equation}
is the time constant of spreading in stage I.

The solution of Eq.~(\ref{eq:c41}) is given by
\begin{equation}
\theta=\theta_{0}\left(1+\theta_{0}\frac{t}{\Gamma_{\rm I}}\right)^{-1},
\label{eq:c43}
\end{equation}
where $\theta_{0}$ is the initial contact angle at $t=0$, which gives the power law for the contact angle $\theta$ 
\begin{equation}
\theta \propto t^{-1},
\label{eq:c44}
\end{equation}
and that for the contact-line radius $r_{\rm L}$ given by
\begin{equation}
r_{\rm L} =R\sin\phi \propto \theta \propto t^{-1}
\label{eq:c45}
\end{equation}
from Eqs.~(\ref{eq:c16}) and (\ref{eq:c18}).  Then, the spreading velocity decelerates according to the power law,
\begin{equation}
U  \propto \theta^{2} \propto t^{-2}
\label{eq:c46}
\end{equation}
from Eq.~(\ref{eq:c37}).  The spreading of a free and a Cassie droplet in stage I is characterized by the usual power-law rule~\cite{deGennes1985,Bonn2009,Voinov1976,Tanner1979}.  The spreading exponent does not depend on the power-law exponent $n$ of the non-Newtonian liquids because the dissipation is due to the friction which does not depend on $n$.  In contrast to the spreading on a flat substrate, where the MKT predicts the power-law $\theta\propto t^{-3/7}\sim t^{-0.43}$~\cite{deRuijter2000,Roques-Carmes2010}, the droplet relaxes much faster on a spherical substrate ($\theta\propto t^{-1}$).  This is due to the fact that the contact line shrinks ($r_{\rm L}\propto\theta \rightarrow 0$) from Eq.~(\ref{eq:c31}) on a spherical substrate, while it expands ($r_{\rm L}\propto\theta^{-1/3}\rightarrow \infty$)~\cite{deGennes1985,Bonn2009,Iwamatsu2017c} on a flat substrate.  Then, the frictional dissipation at the contact line is more effective on a spherical substrate than on a flat substrate. 

The energy balance equation for the Wenzel droplet is given by
\begin{equation}
\sigma_{\rm LV}\left(r_{\rm s}\cos\theta_{\rm Y}-1\right)
=\zeta U
\label{eq:c47}
\end{equation}
from Eq.~(\ref{eq:c34}).  The roughness gives a constant driving force for spreading, and the spreading velocity will be constant:
\begin{equation}
U = K_{\rm I}',
\label{eq:c48}
\end{equation}
with 
\begin{equation}
K_{\rm I}'=2\left(r_{\rm s}\cos\theta_{\rm Y}-1\right)K_{\rm I},
\label{eq:c49}
\end{equation}
where $K_{\rm I}$ is defined by Eq.~(\ref{eq:c39}).  

Equation~(\ref{eq:c47}) is written as
\begin{equation}
1=-\Gamma_{\rm I}'\dot{\theta},
\label{eq:c50}
\end{equation}
with
\begin{equation}
\Gamma_{\rm I}'=\frac{\Gamma_{\rm I}}{2\left(r_{\rm s}\cos\theta_{\rm Y}-1\right)}
\label{eq:c51},
\end{equation}
where $\Gamma_{\rm I}$ is defined by Eq.~(\ref{eq:c42}).  The solution of Eq.~(\ref{eq:c50}) is given by
\begin{equation}
\theta = \frac{t_{0}-t}{\Gamma_{\rm I}'},
\label{eq:c52}
\end{equation}
where $t_{0}$ is the time when the spreading is completed.  Equation~(\ref{eq:c52}) gives the asymptotic forms
\begin{equation}
\theta \propto r_{\rm L} \propto t_{0}-t.
\label{eq:c53}
\end{equation}
Then, the contact angle $\theta$ and the contact-line radius $r_{\rm L}$ decrease linearly with time $t$ and the spreading velocity remains a constant in accordance with Eq.~(\ref{eq:c48}).  Therefore, the spreading of the Wenzel droplet does not follow the usual power law, and is much faster than the power-law relaxation of a free and a Cassie droplet given by Eq.~(\ref{eq:c44}).  The spreading on a spherical substrate is qualitatively different from that on a flat substrate.

\subsection{Stage II: $\theta_{\rm ls}>\theta$ ($n\le1/2$), or $\theta_{\rm ls}> \theta >\theta_{\rm vf}$ ($n>1/2$) }

This is the final stage of spreading for shear-thinning droplets when  $n\le1/2$ and is the intermediate stage when $n>1/2$ and  $\theta_{\rm ls}>\theta_{\rm vf}$. The main driving force is the line-tension force $f_{\rm l}$ in Eq.~(\ref{eq:c20}) and the main dissipation channel remains the frictional dissipation $\dot{\Sigma}_{\rm f}$ in Eq.~(\ref{eq:c27}).

However, this stage II will appear only when $\theta_{\rm ls}>\theta_{\rm vf}$.  This inequality may not be satisfied as the critical contact angle $\theta_{\rm ls}$ depends on the magnitude of the scaled line tension $\tilde{\tau}$ and the radius $R$ of the spherical substrate and the cavity so that it would be very small compared to the macroscopic critical angle $\theta_{\rm vf}$ ($\theta_{\rm vf}\ll \theta_{\rm ls}$).   In such a case, the spreading consists of three stages I$\rightarrow$III$\rightarrow$IV when $n>1/2$.

Now, the energy balance equation in stage II is given by
\begin{equation}
\sigma_{\rm LV}\left(\left|\frac{r_{0}-R}{r_{0}}\right|\frac{\tilde{\tau}}{\theta}\right)
=\zeta U
\label{eq:c54}
\end{equation}
from Eqs.~(\ref{eq:c33}) or (\ref{eq:c34}).  This equation is valid for all types of droplets (free, Cassie and Wenzel).  Equation (\ref{eq:c54}) predicts
\begin{equation}
U =K_{\rm II} \theta^{-1},  
\label{eq:c55}
\end{equation}
with 
\begin{equation}
K_{\rm II}=\frac{\tau}{\zeta R}\left|\frac{r_{0}-R}{r_{0}}\right|,
\label{eq:c56}
\end{equation}
where we have used the definition in Eq.~(\ref{eq:c8}).  The spreading velocity $U$ accelerates $U\rightarrow \infty$ as $\theta\rightarrow 0$.

Using Eq.~(\ref{eq:c40}), Eq.~(\ref{eq:c54}) is written as
\begin{equation}
\theta^{-1}
=\left(-\Gamma_{\rm II}\dot{\theta}\right),
\label{eq:c57} 
\end{equation}
where
\begin{equation}
\Gamma_{\rm II}=\frac{\zeta}{\tau}\left(\frac{r_{0}}{r_{0}-R}R\right)^{2}
\label{eq:c58}
\end{equation}
is the time constant of stage II.  The solution of Eq.~(\ref{eq:c57}) is given by
\begin{equation}
\theta = \left(\frac{2\left(t_{0}-t\right)}{\Gamma_{\rm II}}\right)^{\frac{1}{2}}.
\label{eq:c59}
\end{equation}
Then, the spreading finishes within a finite time $t_{0}$ according to
\begin{equation}
\theta \propto r_{\rm L} \propto \left(t_{0}-t\right)^{\frac{1}{2}}.
\label{eq:c60}
\end{equation}
The time evolution of the spreading velocity $U$ is easily obtained from Eqs.~(\ref{eq:c55}) and (\ref{eq:c60}).  It accelerates and will diverge at the terminal time ($t\rightarrow t_{0}$).

\subsection{Stage III: $\theta_{\rm vf}>\theta>\theta_{\rm ls}$ ($n>1/2$) }

In stage III, the main driving force is the surface tension force $f_{\rm s}$ in Eq.~(\ref{eq:c2}) and the main dissipation is the viscous dissipation $\dot{\Sigma}_{\rm v}$  in Eq.~(\ref{eq:c26}) characterized by the viscosity $\kappa$ and the power-law exponent $n$.  Since this stage III for a droplet on a spherical substrate has already been considered in our previous paper~\cite{Iwamatsu2017c}, and the generalization to the droplet within a spherical cavity is straightforward, we will only summarize the final results.  

The energy balance equation for the free and the Cassie droplet is given by
\begin{equation}
\sigma_{\rm LV}\frac{1}{2}\theta^{2}
=\lambda\left(\frac{2n+1}{n}\right)^{n}\left|\frac{r_{0}}{r_{0}-R}R\right|^{1-n}\kappa U^{n}\theta^{1-2n}
\label{eq:c61}
\end{equation}
from Eq.~(\ref{eq:c33}), which gives
\begin{equation}
U^{n} = K_{\rm III}\theta^{2n+1},
\label{eq:c62}
\end{equation}
where the proportionality constant is given by
\begin{equation}
K_{\rm III}=\frac{\sigma_{\rm LV}}{2\kappa\lambda}\left(\frac{2n+1}{n}\right)^{-n}\left|\frac{r_{0}}{r_{0}-R}R\right|^{n-1}.
\label{eq:c63}
\end{equation}

Using Eq.~(\ref{eq:c40}), Eq.~(\ref{eq:c61}) is simplified to
\begin{equation}
\theta^{2n+1}
=\left(-\Gamma_{\rm III}\dot{\theta}\right)^{n},
\label{eq:c64} 
\end{equation}
where
\begin{equation}
\Gamma_{\rm III}=\left(\frac{2n+1}{n}\right)\left[\frac{2\kappa\lambda}{\sigma_{\rm LV}}\left|\frac{r_{0}}{r_{0}-R}R\right|\right]^{\frac{1}{n}}
\label{eq:c65}
\end{equation}
is the time constant of stage III~\cite{Iwamatsu2017c}.  The solution of Eq.~(\ref{eq:c64}) is given by
\begin{equation}
\theta = \theta_{0}\left(1+\frac{n+1}{n}\theta_{0}^{\frac{n+1}{n}}\frac{t}{\Gamma_{\rm III}}\right)^{-\frac{n}{n+1}},
\label{eq:c66}
\end{equation}
which gives the power law
\begin{equation}
\theta \propto r_{\rm L} \propto t^{-\frac{n}{n+1}},
\label{eq:c67}
\end{equation}
and a similar power law for the spreading velocity $U$ from Eqs.~(\ref{eq:c62}) and (\ref{eq:c67}).  The spreading of Newtonian liquid ($n=1$) on a spherical substrate, for example, is characterized by the power law $\theta\propto t^{-1/2}$ from Eq.~(\ref{eq:c67}).  This result is different from the famous Tanner's law $\theta\propto t^{-3/10}\sim t^{-0.33}$ on a flat substrate~\cite{Tanner1979,deRuijter2000}, which corresponds to the spreading law of stage III on a flat substrate.

The energy balance equation for the Wenzel droplet is given by
\begin{equation}
\sigma_{\rm LV}\left(r_{\rm s}\cos\theta_{\rm Y}-1\right)
=\lambda\left(\frac{2n+1}{n}\right)^{n}\left|\frac{r_{0}}{r_{0}-R}R\right|^{1-n}\kappa U^{n}\theta^{1-2n}
\label{eq:c68}
\end{equation}
from Eq.~(\ref{eq:c34}), which gives
\begin{equation}
U^{n}=K_{\rm III}'\theta^{2n-1},
\label{eq:c69}
\end{equation}
where
\begin{equation}
K_{\rm III}' = 2\left(r_{\rm s}\cos\theta_{\rm Y}-1\right)K_{\rm III},
\label{eq:c70}
\end{equation}
and the evolution equation of the contact angle $\theta$ becomes
\begin{equation}
\theta^{2n-1}
=\left(-\Gamma_{\rm III}'\dot{\theta}\right)^{n},
\label{eq:c71}
\end{equation}
with
\begin{equation}
\Gamma_{\rm III}'=\frac{\Gamma_{\rm III}}{2^{\frac{1}{n}} \left(r_{\rm s}\cos\theta_{\rm Y}-1\right)^{\frac{1}{n}}},
\label{eq:c72}
\end{equation}
whose solutions are different for shear-thickening liquids with $n>1$, shear-thinning liquids with $n<1$, and  Newtonian liquids with $n=1$~\cite{Iwamatsu2017c}.  They are summarized as
\begin{equation}
\theta =
\begin{cases}
\theta_{0}\left(1+\frac{n-1}{n}\theta_{0}^{\frac{n-1}{n}}\frac{t}{\Gamma_{\rm III}'}\right)^{-\frac{n}{n-1}} & n>1, \\
\theta_{0}\exp\left(-\frac{t}{\Gamma_{\rm III}'}\right) & n=1, \\
(\frac{1-n}{n}\frac{t_{0}-t}{\Gamma_{\rm III}'})^{\frac{n}{1-n}} & n<1,
\end{cases}
\label{eq:c73}
\end{equation}
whose asymptotic form becomes~\cite{Iwamatsu2017c}
\begin{equation}
\theta \propto r_{\rm L} \propto
\begin{cases}
t^{-\frac{n}{n-1}} & n>1, \\
\exp\left(-\frac{t}{\Gamma_{\rm III}'}\right) & n=1, \\
\left(t_{0}-t\right)^{\frac{n}{1-n}} & n<1.
\end{cases}
\label{eq:c74}
\end{equation}
The corresponding asymptotic form of the spreading velocity $U$ can be easily derived from Eqs.~(\ref{eq:c69}) and (\ref{eq:c74}), and is also classified into three types.  The power-law relaxation of the contact angle $\theta$ for the shear-thickening liquids ($n>1$), the exponential relaxation for the Newtonian liquid ($n=1$), and the faster spreading within a finite time $t_{0}$ for the shear-thinning liquids ($n<1$) are predicted for the Wenzel droplet on a spherical substrate~\cite{Iwamatsu2017c}.

\subsection{Stage IV: $\theta<\theta_{\rm ls}$ and $\theta<\theta_{\rm vf}$ ($n>1/2$) }

In the final stage of spreading, when $n>1/2$, the dynamic contact angle $\theta$ becomes low so that the inequalities $\theta<\theta_{\rm ls}$ and $\theta<\theta_{\rm vf}$ hold simultaneously.  Then, the main driving force is the line-tension force $f_{\rm l}$ in Eq.~(\ref{eq:c20}) and the main dissipation is the viscous dissipation $\dot{\Sigma}_{\rm v}$ in Eq.~(\ref{eq:c29}).  The energy balance equation becomes
\begin{equation}
\sigma_{\rm LV}\left(\left|\frac{r_{0}-R}{r_{0}}\right|\frac{\tilde{\tau}}{\theta}\right)
=\lambda\left(\frac{2n+1}{n}\right)^{n}\left|\frac{r_{0}}{r_{0}-R}R\right|^{1-n}\kappa U^{n}\theta^{1-2n}
\label{eq:c75}
\end{equation}
from Eqs.~(\ref{eq:c33}) and (\ref{eq:c34}) for the free and the Cassie droplet as well as for the Wenzel droplet.  Equation~(\ref{eq:c75}) predicts a relation between the spreading velocity $U$ and the dynamic contact angle $\theta$ given by
\begin{equation}
U^{n} = K_{\rm IV} \theta^{2n-2},
\label{eq:c76}
\end{equation}
with
\begin{equation}
K_{\rm IV}=\frac{\tau}{\kappa\lambda}\left(\frac{2n+1}{n}\right)^{-n}\left|\frac{r_{0}}{r_{0}-R}R\right|^{n-2}.
\label{eq:c77}
\end{equation}
Then, the spreading velocity decelerates as $\theta\rightarrow 0$ for shear-thickening liquids ($n>1$), while it accelerates for the shear-thinning liquids ($n<1$).  The spreading velocity remains constant for Newtonian liquids ($n=1$).

Using Eq.~(\ref{eq:c40}), Eq.~(\ref{eq:c75}) is simplified to
\begin{equation}
\theta^{2n-2}
=\left(-\Gamma_{\rm IV}\dot{\theta}\right)^{n},
\label{eq:c78} 
\end{equation}
with
\begin{equation}
\Gamma_{\rm IV}=\left(\frac{2n+1}{n}\right)\left[\frac{\kappa\lambda}{\tau}\left(\frac{r_{0}}{r_{0}-R}R\right)^{2}\right]^{\frac{1}{n}},
\label{eq:c79}
\end{equation}
which determines the time scale of spreading for the free droplet on a completely wettable smooth substrate and the Cassie and the Wenzel droplet on a completely wettable rough substrate.  Note that the time scale $\Gamma_{\rm IV}$ does not depend on the roughness $r_{\rm s}$ of the substrate.  It depends on the viscosity $\kappa$, the line tension $\tau$, the power-law exponent $n$, and the geometric parameters of the radius of the substrate $R$ and the volume of the droplet $V_{0}$ through $r_{0}$. The time evolution equation in Eq.~(\ref{eq:c78}) is the same as that derived for the free droplet on a spherical smooth substrate when the line-tension effect is dominant~\cite{Iwamatsu2017b}.  Here, Eq.~(\ref{eq:c78}) describes the evolution of the droplet not only on a convex spherical substrate but also in a concave spherical cavity.

The solution of Eq.~(\ref{eq:c78}) is summarized as
\begin{equation}
\theta =
\begin{cases}
\theta_{0}\left(1+\frac{n-2}{n}\theta_{0}^{\frac{n-2}{n}}\frac{t}{\Gamma_{\rm IV}}\right)^{-\frac{n}{n-2}} & n>2, \\
\theta_{0}\exp\left(-\frac{t}{\Gamma_{\rm IV}}\right) & n=2, \\
\left(\frac{2-n}{n}\frac{t_{0}-t}{\Gamma_{\rm IV}}\right)^{\frac{n}{2-n}} & n<2,
\end{cases}
\label{eq:c80}
\end{equation}
which gives the asymptotic forms
\begin{equation}
\theta \propto r_{\rm L} \propto
\begin{cases}
t^{-\frac{n}{n-2}} & n>2, \\
\exp\left(-\frac{t}{\Gamma_{\rm IV}}\right) & n=2, \\
\left(t_{0}-t\right)^{\frac{n}{2-n}} & n<2,
\end{cases}
\label{eq:c81}
\end{equation}
and the similar three types of spreading velocity $U$ from Eqs.~(\ref{eq:c76}) and (\ref{eq:c81}). The contact angle $\theta$ and the radius $r_{\rm L}$ change linearly with time $t$ for Newtonian liquids ($n=1$).  This stage IV for the droplet on a rough spherical substrate of the Cassie and the Wenzel droplets are the same as that for the free droplet on a smooth spherical substrate~\cite{Iwamatsu2017b}.

So far, we have assumed that the line tension $\tau$ is positive and constant.  However, the line tension cannot be constant in the last stage of spreading when $\theta\rightarrow 0$ since the effective equilibrium contact angle $\theta_{\rm e}'$ cannot be defined since the line-tension contribution in Eq.~(\ref{eq:c7}) diverges as $\theta\rightarrow 0$ or $\phi\rightarrow \pi/2$.  Similarly, the line-tension force $f_{\rm l}$ in Eq.~(\ref{eq:c20}) diverges as the radius of contact line shrinks ($r_{\rm L}\rightarrow 0$) and the contact angle vanishes ($\theta\rightarrow 0$).

This problem is partially circumvented, for example, by assuming that the line tension $\tau$ depends on the contact angle $\theta$~\cite{Marmur1997,Herminghaus2006,Law2017,Kanduc2017,Iwamatsu2018} through
\begin{equation}
\tau=\tau_{0}\theta^{\beta},
\label{eq:c82}
\end{equation}
where $\tau_{0}$ is a constant and $\beta>1$ is another exponent which characterizes the contact-angle dependence of line tension $\tau$.  For example, $\beta=2$ was derived for the gravitation contribution to the line tension by Herminghaus and Brochard~\cite{Herminghaus2006}.  Then, the divergence of the left-hand side of Eq.~(\ref{eq:c75}) can be avoided as far as $\beta>1$, and Eqs.~(\ref{eq:c76}) and (\ref{eq:c78}) are modified to
\begin{eqnarray}
U^{n} &=& K_{\rm IV}' \theta^{2n-2+\beta}, 
\label{eq:c83} \\
\theta^{2n-2+\beta}
&=&\left(-\Gamma_{\rm IV}'\dot{\theta}\right)^{n},
\label{eq:c84} 
\end{eqnarray}
where $K_{\rm IV}'$ and $\Gamma_{\rm IV}'$ are those given by Eq.~(\ref{eq:c77}) and Eq.~(\ref{eq:c79}) with $\tau$ replaced by $\tau_{0}$ in Eq.~(\ref{eq:c82}).  The asymptotic forms in Eq.~(\ref{eq:c80}) are now modified to
\begin{equation}
\theta \propto r_{\rm L} \propto
\begin{cases}
t^{-\frac{n}{n-2+\beta}} & n>2-\beta, \\
\exp\left(-\frac{t}{\Gamma_{\rm IV}'}\right) & n=2-\beta, \\
\left(t_{0}-t\right)^{\frac{n}{2-n-\beta}} & n<2-\beta,
\end{cases}
\label{eq:c85}
\end{equation}
so that the exponents are also modified.  A similar modification is possible for stage II, which can be the last stage of spreading ($\theta\rightarrow 0$) when $n<1/2$.  Then, we have three types of asymptotic forms similar to Eqs.~(\ref{eq:c85}) instead of Eq.~(\ref{eq:c60}) depending on the magnitude of the exponent $\beta$.

Mathematically, Eq. (\ref{eq:c82}) is the simplest way to avoid divergence of the line-tension effect when $\theta\rightarrow 0$.  Physically, however, the existence of this stage IV when $\theta<\theta_{\rm ls}$ and $\theta<\theta_{\rm vf}$ itself is a delicate problem as the scaled line tension $\tilde{\tau}$ can be very small and, therefore, $\theta_{\rm ls}$ defined by Eqs.~(\ref{eq:c21}) and (\ref{eq:c22}) can be a microscopic atomic scale.  Then, our formulation based on the energy balance approach in Sec. II would break down because the atomic scale cut off distance must always be introduced to calculate viscous dissipation~\cite{deGennes1985,Brochard-Wyart1992,Snoeijer2013,Iwamatsu2017a,Liang2012} and the inequality $\theta<\theta_{\rm ls}$ cannot be satisfied. Therefore, the existence of this stage IV as well as stage II, which also requires the inequality $\theta<\theta_{\rm ls}$, depends on the size of the scaled line tension $\tilde{\tau}$.  Even if these stages II and IV exist, their experimental observation will not be easy as the size scale will be nanoscale.

\subsection{Spreading on a spherical substrate and in a spherical cavity }

In Table~\ref{tab:T1}, we summarize the exponent $\alpha$ which characterizes the relation between the spreading velocity $U$ and the contact angle $\theta$ through $U\propto \theta^{\alpha}$ for four stages of spreading I-IV.  The spreading of the droplet follow the stage I$\rightarrow$II when $n<1/2$ and I$\rightarrow$II$\rightarrow$IV if $\theta_{\rm ls}>\theta_{\rm vf}$ or I$\rightarrow$III$\rightarrow$IV if $\theta_{\rm ls}<\theta_{\rm vf}$ when  $n>1/2$.  Mid-stage II may not appear unless the droplet and the substrate is nanoscale. The exponent $\alpha$ depends on the power-law exponent $n$ of the non-Newtonian liquids only when the main dissipation channel is the viscous dissipation. 

\begin{table}[htb]
 \begin{center}
  \caption{The exponent $\alpha$ defined by $U\propto \theta^{\alpha}$ for the four stages of spreading I-IV}
  \begin{tabular}{c|c|c} 
\hline
Stage & Free and Cassie  & Wenzal  \\
\hline
I &
2 &
0 \\
II &
-1 &
-1 \\
III &
$\frac{2n+1}{n}$ &
$\frac{2n-1}{n}$ \\
IV &
$\frac{2n-2}{n}$ &
$\frac{2n-2}{n}$ \\
\hline
  \end{tabular}
  \label{tab:T1}
 \end{center}
\end{table}

The time evolution of the contact angle $\theta$ is summarized in Table~\ref{tab:T2}.  A free droplet and a Cassie droplet of Newtonian liquids ($n=1$), for example, evolves $\theta\propto t^{-1}$ in early stage I, then it follows $\theta\propto \sqrt{t_{0}-t}$ in statge II or $\theta\propto t^{-1/2}$ in stage III and finally $\theta\propto \left(t_{0}-t\right)$ to accomplish complete wetting in stage IV.  A popular power-law spreading similar to the famous Tanner's low appears only in early stages I and III though the exponent is $1$ and $1/2$ instead of $3/10$ of the Tanner's law~\cite{Tanner1979} or $3/7$ of the molecular kinetic theory~\cite{deRuijter2000} on a flat substrate.  Therefore, the spreading on a convex and a concave spherical substrate will be generally faster than the spreading on a flat substrate~\cite{Iwamatsu2017c}.

\begin{table}[htb]
 \begin{center}
  \caption{The time evolution of the contact angle $\theta$ for the four stages of spreading I to IV.}
  \begin{tabular}{c|cc|cc} 
\hline
Stage & Free and Cassie  & & Wenzel & \\
\hline
I &
 $t^{-1}$ &
&
 $\left(t_{0}-t\right)$ &
\\
II &
 $\left(t_{0}-t\right)^{\frac{1}{2}}$ &
 &
 $\left(t_{0}-t\right)^{\frac{1}{2}}$ &
 \\
III &
 $t^{-\frac{n}{n+1}}$ &
 &
 $t^{-\frac{n}{n-1}}$ &
 ($n>1$) \\
&
&
&
$\exp\left(\frac{t}{\Gamma_{\rm III}'}\right)$ &
($n=1$)\\
&
&
&
$\left(r_{0}-t\right)^{\frac{n}{1-n}}$ &
($n<1$)\\
IV &
$t^{-\frac{n}{n-2}}$ &
($n>2$) &
$t^{-\frac{n}{n-2}}$ &
($n>2$) \\
&
$\exp\left(-\frac{t}{\Gamma_{\rm IV}}\right)$ &
($n=2$) &
$\exp\left(-\frac{t}{\Gamma_{\rm IV}}\right)$ &
($n=2$) \\
&
$\left(t_{0}-t\right)^{\frac{n}{2-n}}$ &
($n<2$) &
$\left(t_{0}-t\right)^{\frac{n}{2-n}}$ &
($n<2$) \\
\hline
  \end{tabular}
  \label{tab:T2}
 \end{center}
\end{table}

The spreading of a Wenzel droplet of Newtonian liquids ($n=1$), for example, is faster than the free and the Cassie droplet since the constant capillary force $\sigma_{\rm LV}\left(r_{\rm s}\cos\theta_{\rm Y}-1\right)$ in Eq.~(\ref{eq:c15}) always accelerates the spreading.  The contact angle of the Wenzel droplet of Newtonian liquid ($n=1$) decreases linearly with time $\theta\propto \left(t_{0}-t\right)$ in early stage I (Table~\ref{tab:T2}).  Subsequently, it decreases as $\theta\propto\sqrt{t_{0}-t}$ in stage II or exponentially $\theta\propto \exp\left(-t/\Gamma_{\rm III}'\right)$ in stage III.  Finally, the contact angle relaxes linearly $\theta\propto \left(t_{0}-t\right)$ to $\theta\rightarrow 0$ in final stage IV.  Not only the topology (spherical symmetry) but also the topography (roughness) accelerate~\cite{Iwamatsu2017c} the spreading of the Wenzel droplet on convex and concave spherical substrates.  The spreading scenario of non-Newtonian liquid ($n\neq 1$) is more diverse in Table~\ref{tab:T2}.

\begin{figure}[htbp]
\begin{center}
\includegraphics[width=0.9\linewidth]{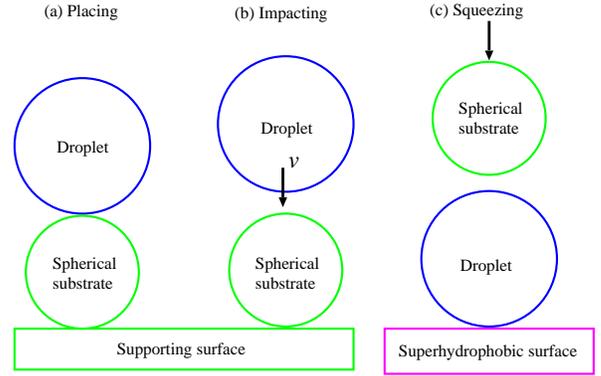}
\caption{
Three ways to realize spreading on a completely wettable spherical substrate.  (a) Placing a spherical droplet quietly on top of the spherical substrate supported by a flat surface.  (b) Impacting a spherical droplet onto a spherical substrate by low velocity $v$.  (c) Squeezing a spherical substrate into a spherical droplet supported by a superhydrophobic surface.  
 }
\label{fig:C5}
\end{center}
\end{figure}

There exists a wealth of experimental data of spreading of Newtonian as well as non-Newtonian liquids on a flat smooth substrate~\cite{Bonn2009,Liang2012}.   There also exists a small number of experimental data on a flat rough substrate~\cite{Apel-Paz1999,McHale2004,Xu2008,McHale2009}.  However, the problem of spreading on a spherical substrate has attracted almost no attention so far except for few experimental studies of wetting on a smooth~\cite{Tao2011,Guilizzoni2011,Eral2011,Extrand2012,Wu2015} and a rough spherical substrate~\cite{Byon2010}.

The experimental verification of the mathematically simple spreading laws summarized in Table ~\ref{tab:T2} will not be easy for the stages II and IV as the line tension is the main driving force and the nanometer scale droplets and substrates will be required.  However, the observation of stages I and III will be possible as the surface tension is the main driving force and the macroscopic droplets and substrates can be used.  In fact, there are many experimental evidences that the spreading process on flat substrates can be divided into several stages~\cite{deRuijter1999,Roques-Carmes2010,Bazazi2018,James2018}.

Our simple spreading laws on a spherical substrate are based on many simplifying assumptions.  The key assumption is that the shape of the droplet can remain spherical, and the contact line can go around the equator of the substrate and cut in the lower hemisphere from the upper hemisphere.  In this late stage of complete wetting, the substrate is almost completely engulfed in the spherical droplet and the spreading is accompanied by the shrinkage of the contact line as shown in Fig.~\ref{fig:C1}.  However, to the best of our knowledge, most of the previous experimental studies of wetting and spreading on a spherical substrate are concentrated on the droplet, which starts from the north pole, and remains on the upper hemisphere~\cite{Tao2011,Eral2011,Guilizzoni2011,Extrand2012,Wu2015}.  Naturally, the spreading in a cavity is more difficult to study experimentally so that only theoretical or numerical studies exist~\cite{Lefevre2004,Li2008,Bormashenko2013,Iwamatsu2016d,Zhang2016,Tinti2017}.

Recently, a droplet impacts onto a stationary solid sphere has been studied intensively~\cite{Bakshi2007,Mitra2013,Banitabaei2017}.  Although most of the studies were done for the hydrophobic spheres, there are a limited number of experimental data for a hydrophilic sphere~\cite{Banitabaei2017}.  Since those studies pay most attention to the morphology of the droplet, which depends on the impact velocity, wettability of sphere, and the size ratio of the droplet and sphere, no quantitative data of spreading such as the spreading velocity or the time dependence of the contact angle was reported.  However, the morphology of the late stage of spreading on a hydrophilic spherical substrate~\cite{Banitabaei2017} resembles the scenario shown in Fig.~\ref{fig:C1}(a). Therefore, it will be possible to confirm our spreading laws of stage I and III in the near future.  To this end, the late stage configuration in Fig.~\ref{fig:C1}(a) can be realized not only by placing or impacting a spherical droplet on a stationary sphere [Fig.~\ref{fig:C5}(a) and 5(b)] but also by squeezing a spherical substrate into a spherical droplet supported by a super hydrophobic flat surface~\cite{Biance2004} [Fig.~\ref{fig:C5}(c)].

Although we can derive simple spreading laws even on rough textured spherical substrates (Fig.~\ref{fig:C3}) by adopting the Cassie and the Wenzel model~\cite{Shuttleworth1948,McHale2004,McHale2009,Wenzel1936,Bormashenko2015,Cassie1944}, the experimental observation of those scaling laws will be more difficult on rough spherical substrates than on smooth spherical substrates.  In fact, all experimental studies so far are concerned with the early stage of spreading on rough flat substrates because the energy dissipation due to imbibition~\cite{Ishino2007,Grewal2015} will be negligible only in the early stage~\cite{Xu2008,Kim2013,Zong2018}.  Further efforts are certainly necessary to understand and unify the spreading and the imbibition not only on spherical substrates on flat substrates as well.

\section{\label{sec:sec4} Conclusion}

In the present study, we considered the problem of spreading of a droplet of non-Newtonian power-law liquids on a convex spherical substrate and in a concave spherical cavity using the energy balance approach.  Not only a smooth surface but also a rough surface is considered.  The Wenzel~\cite{Wenzel1936} and the Cassie-Baxter~\cite{Cassie1944} model are adopted on a rough substrate.  Two driving forces of spreading by the surface tension (capillary) force and the line-tension force are considered.  The latter becomes important only on a spherical substrate as the contact-line radius shrinks.  Also, two dissipation channels due to the viscous dissipation within the bulk and the frictional dissipation at the contact line are considered.  The former is the main dissipation channel of the hydrodynamic theory~\cite{deGennes1985,Brochard-Wyart1992,Bonn2009,Voinov1976,Tanner1979} and the latter is the main dissipation of the molecular kinetic theory~\cite{Blake1969,deRuijter1999b,Blake2006,Bertrand2009}.   Therefore, we extended the combined theory of spreading~\cite{deRuijter1999,deRuijter1999b,Roques-Carmes2010} on flat substrates to that on convex and concave spherical substrates.  Furthermore, our theory includes the line tension as the driving force of spreading.

The time evolution of the contact angle can be classified into four stages.  In each stage, the dynamic contact angle has a characteristic time dependence. The time evolution can be a power-law relaxation, an exponential relaxation, and a relaxation within a finite time.   Even when the time evolution is standard power-law relaxation, the exponent is different from those derived on a flat substrate due to the spherical symmetry and the shrinking contact line.  Generally speaking, the spreading on a spherical substrate is faster than that on a flat substrate.  The spreading of the Cassie droplet on a spherical rough substrate is the same as that of the free droplet on a smooth substrate.  However, the spreading of the Wenzel droplet is faster than that of the Cassie-Baxter model on a spherical rough substrate.  All those results summarized in Tables~\ref{tab:T1} and \ref{tab:T2} reveal the diversity of  {\it topography and topology driven spreading} on a spherical substrate~\cite{Iwamatsu2017c} again even when the line-tension force and the contact-line friction are included.

The experimental as well as numerical verification of our theoretical predictions are highly desired and urgent.  The theoretical results presented in this paper are merely a first step toward the complete understanding of spreading on various complex curved substrates. Our theoretical predictions together with those future experimental and numerical results would be valuable to design and develop new nano materials, nano devices, and new engineering applications.






\end{document}